\documentclass[12pt]{article}
\usepackage[cp1251]{inputenc}
\usepackage[english]{babel}
\usepackage{graphicx}
\usepackage{amsmath}
\usepackage{epsfig}
\renewcommand{\abstract}{}

\begin{document}
\fontsize{12}{12} \selectfont
\title{Luminosity Dependence of the Quasar Clustering from SDSS DR5}
\author{Ganna Ivashchenko}
\date{}
\maketitle
\begin{center} {\small Kyiv National Taras Shevchenko University, Glushkova ave., 2, 03127 Kyiv, Ukraine\\ hanna@univ.kiev.ua}
\end{center}

\begin{abstract}
43024 objects, which were primarily identified as quasars in SDSS
DR5 and have spectroscopic redshifts were used to study the
luminosity dependence of the quasar clustering with the help of two
different techniques. The obtained results reveal that brighter
quasars are more clustered, but this dependence is weak, which is in
agreement with the results by Porciani \& Norberg
\cite{porciani2006} and theoretical predictions by Lidz et al.
\cite{lidz}.


\end{abstract}

\section{Introduction}

\indent \indent Determining the distribution of extragalactic
objects is one of the most important problems of modern cosmology,
because they are the only tracers of the dark matter, except
anisotropy of the CMB, that helps us to realize the matter
distribution on the redshifts of about 1000. Unfortunately even in
the local Universe it is difficult to see the most faint galaxies,
and when we go farther to larger redshifts, the most part of objects
we can observe there with the modern ground-based telescopes used
for large surveys are quasars as the most luminous objects. The
matter distribution that we could reconstruct with the help of
quasars is just like a light sketch, but it is all we have for
today.

The quasars are nonuniform objects: they have various luminosities
and are on the different stages of their evolution. Thus the
reasonable question arises: how their clustering could depend on
their physical properties, for example on their luminosity?
According to different numerical simulations of galaxy mergers that
incorporate black hole growth, this dependence has to exist because
more luminous quasars are considered to be born in denser
environment. In the main part of such models, in which the host halo
mass correlates with the instantaneous luminosity of the quasars
(see e.g. \cite{kaufman}), there should be a strong luminosity
dependence of the quasar clustering. In the other type of such
models, in which the host halo mass correlates with the peak
luminosity of quasars, the luminosity dependence of the quasar
clustering should be weaker because all of the quasars we see now,
are considered to be similar objects but on different stages of
their evolution (see \cite{lidz} and references therein). It is
worth to note, that the redshift distribution of the quasars is not
the same for different luminosities due to possible evolution
effects.

The largest quasar surveys are 2dF and SDSS. In contrast to 2dF,
which is finished for today and has 2QZ catalogue as a result
\cite{2dF-XII}, SDSS \cite{SDSS5} is in progress and the area
covered by it is increasing. Only a part of objects primarily
classified as quasars were justified by spectra analysis and were
included into SDSS Quasar Catalogue IV \cite{dr4}. However even
photometric classification of quasars with color diagrams
\cite{phot} is sufficient for using these objects for statistical
purposes \cite{myers2, myers1}.

Some attempts to find luminosity dependence of the quasar clustering
have been made with different samples. E.g. Adelberg \& Steidel
\cite{adelberg}, who worked with their own survey, pointed to
luminosity independent quasar clustering. 2dF-team, that studied 2QZ
survey and found the little redshift evolution in the amplitude of
the power spectrum \cite{2dF-IV}, \cite{2dF-XI} and significant
increase in clustering amplitude at high redshifts \cite{2dF-XIV,
2dF-II}, detected only marginal evidence for quasars with brighter
apparent magnitudes to have a stronger clustering amplitude
\cite{2dF-IX}. Porciani and Norberg \cite{porciani2006} also found
weak luminosity dependence of the clustering in 2QZ survey.
Furthermore they noted that samples with different redshifts show
different trends in luminosity dependence: in the redshift bin
$1.7<z<2.1$ the brightest quasars seem to be more clustered (have
larger bias parameter); in the redshift bin $1.3<z<1.7$ the bias
parameter seems to follow a U-shape; and the low redshift bin
($0.8<z<1.3$) does not show any particular trend. Myers et al.
\cite{myers2}, working with $\sim$300,000 photometrically classified
quasars from the 4th Data Release of the SDSS, detected no
significant luminosity dependence and pointed out that a 3$\sigma$
detection of such effect requires a sample several times larger.

The present work deals with the study of the luminosity dependence
of the quasar clustering for quasars from the Fifth Data Release of
SDSS. The sample and its peculiarities are described in Section 2.
The results of estimation of the correlation length as a function of
luminosity are presented in Section 3. As we do not have such large
sample, which we need, according to Myers et al. \cite{myers2}, for
precise measurements of the correlation function, we used another
technique, which is described in Section 4. Finally, Section 5 is
dedicated to discussion of obtained results.

\section{The Data}

\begin{table}[!b]\centering
\begin{minipage}[b]{.80\linewidth}
\centering \caption{The samples used for both methods. N is the
number of objects for all given redshift interval ($z$). $M_{g,eff}$
is effective absolute magnitude, $N_{tot}$ is the number of objects
with luminosity within given range ($M_{g}$) and $\bar{z}$ is the mean redshift.}
\medskip
\begin{tabular}{|c|c|c|c|c|c|}
\hline
$z$ & $N$ & $M_{g}$ & $M_{g,eff}$ & $N_{tot}$ & $\bar{z}$ \\
\hline
 	       &      & $-24.5\div-23.5$ & -24.1 & 2209 & $0.937$ \\
$0.80\div1.10$ & 8621 & $-25.5\div-24.5$ & -25.0 & 4564 & $0.965$\\
 	       &      & $-26.5\div-25.5$ & -25.9 & 1202 & $0.984$\\
\hline
 	       &      & $-25.0\div-24.0$ & -24.6 & 2065 & $1.213$\\
$1.10\div1.35$ & 8619 & $-26.0\div-25.0$ & -25.5 & 5016 & $1.230$\\
 	       &      & $-27.0\div-26.0$ & -26.3 & 1210 & $1.244$\\
\hline
 	       &      & $-25.5\div-24.5$ & -25.1 & 2339 & $1.466$\\
$1.35\div1.59$ & 8579 & $-26.5\div-25.5$ & -25.9 & 4859 & $1.480$\\
 	       &      & $-27.5\div-26.5$ & -26.8 & 1105 & $1.490$\\
\hline
 	       &      & $-25.5\div-24.5$ & -25.2 & 1613 & $1.702$\\
$1.59\div1.83$ & 8511 & $-26.5\div-25.5$ & -26.1 & 4520 & $1.704$\\
 	       &      & $-27.5\div-26.5$ & -26.9 & 2061 & $1.719$\\
\hline
	       &      & $-26.0\div-25.0$ & -25.6 & 2316 & $1.968$\\
$1.83\div2.20$ & 8694 & $-27.0\div-26.0$ & -26.5 & 4377 & $1.977$\\
 	       &      & $-28.0\div-27.0$ & -27.3 & 1714 & $2.019$\\
 \hline
\end{tabular}
\end{minipage}\hfill
\end{table}

\indent \indent The sample of 43024 quasars with spectroscopic
redshifts $0.8<z<2.2$ taken from the 5th Data Release
(http://www.sdss.org/dr5/products/spectra/getspectra.html) of the
Sloan Digital Sky Survey was uses to study the luminosity dependence
of the quasar clustering. The calibrated apparent magnitudes in five
SDSS photometric bands (u,g,r,i,z) are given not as a conventional
Pogson astronomical magnitudes, but as asinh magnitudes
\cite{lupton}. Thus we firstly converted them to Pogson magnitudes.
Then the absolute magnitudes in g-band ($M_{g}$) (the average
wavelength $\lambda_{g}=4686^{\circ}$, magnitude limit
$m_{g,lim}=22.2$) were calculated within the frame of
$\Lambda$CDM-model with $\Omega_{tot}=1$ (spatially flat Universe),
$\Omega_{M}=0.29$, $h=0.73$ \cite{wmap}.

The redshift distribution of quasars in the SDSS survey has several
peaks and valleys, which could not be entirely explained by
different selection effects, like similarity in colors of quasars
within some redshift ranges or presence of some strong emission
lines in different SDSS filter bands (\cite{bell} and references
therein). But in spite of such peculiarities the whole distribution
reveals a large hump on redshift of about 2 which agrees with an
idea, according to which the most part of quasars had to be born at
that time.

To study the luminosity dependence of the quasar clustering one
should choose intervals of redshift small enough to avoid effects of
redshift evolution (see e.g. \cite{2dF-XIV}, \cite{2dF-II},
\cite{myers2}, \cite{myers1}). Therefore our sample was divided into
5 redshift intervals (see Table 1) with the similar number of
quasars. The whole sample covers the so-called SDSS 'window' (the
redshift interval of SDSS data, where the photometrical selected
quasars are the most uncontaminated due to specific character of the
photometrical selection technique). Then in each redshift interval
we selected 3 subsamples according to the absolute magnitude in
g-band -- 'bright', 'medium' and 'faint' quasars. We chose absolute
magnitude intervals of the same size ($\Delta M_{g}=1$).

\section{Correlation functions}

\indent \indent The first technique applied in this work to study
the luminosity dependence of the quasar clustering is calculation of
the redshift-space correlation function for objects with different
luminosities. This means that the cross-correlation between quasars
with luminosity within given range and quasars with any luminosity
was calculated. Both in this case and in the second method, the
measured distances are comoving distances in the reference frame of
the local observer related to the first quasar. All the distances
were measured within the frame of the $\Lambda$CDM-model and
spatially flat Universe with the following parameters:
$\Omega_{M}=0.29$ and $h=0.73$ \cite{wmap}.

According to \cite{peebles} the probability to find a neighbour for
i-th quasar in a spherical layer $[r,r+\Delta r]$ in the epoch $t$
is determined in terms of the two-point correlation function
$\xi(r)$. Note, that all the calculations in this and the fourth
sections were made in redshift-space, as we cannot separate
cosmological redshift and peculiar velocities of quasars. Thus the
total number of neighbours from the whole sample is
\begin{equation*}\label{deltaN}
    \Delta N(r)=4\pi\sum_{i}n(t_{i})\left[\frac{1}{3}[(r+\Delta
    r)^{3}-r^{3}]+\int_{r}^{r+\Delta
    r}\xi(\rho)\rho^{2}d\rho\right],
\end{equation*}
where $n(t_{i})$ is a mean number density of objects in the
neighbourhood of i-th quasar. The similar estimation for random
catalogue, which is considered to represent random spatial
distribution of objects with no clustering, is
\begin{equation*}\label{DeltaNast}
    \Delta N^{\ast}(r)=\frac{4\pi}{3}\sum_{i}n'(t_{i})[(r+\Delta
    r)^{3}-r^{3}].
\end{equation*}
Assuming
\begin{equation*}
    \sum_{i}n'(t_{i})\approx\sum_{i}n(t_{i}),
\end{equation*}
we have
\begin{equation}\label{NN}
    \frac{\Delta N(r)}{\Delta N^{\ast}(r)}-1=\frac{\int_{r}^{r+\Delta r}\xi(\rho)\rho^{2}d\rho}{\Delta r(r^{2}+r\Delta r+\Delta
    r^{2}/3)}.
\end{equation}
The common power-low correlation function was used
\begin{equation} \label{cf}
 \xi(r)=\left(\frac{r_{c}}{r}\right)^{\gamma},
\end{equation}
thus the expression for fitting parameters $r_{c}$, $\gamma$ is the
following
\begin{equation}\label{fit}
    \frac{\Delta N(r)}{\Delta
    N^{\ast}(r)}-1=\frac{r_{c}^{\gamma}}{3-\gamma}\frac{(r+\Delta r)^{3-\gamma}-r^{3-\gamma}}{\Delta r(r^{2}+r\Delta r+\Delta
    r^{2}/3)}.
\end{equation}
As the redshift-space correlation function has some distortions
(like Finger of God effect and $\beta$-distortion) due to peculiar
velocities of the objects (see e.g. \cite{matsubara, ballinger,
peacock}), and they differ on small and large scales, the
correlation function of quasars cannot be approximated with one
power-low function on all the scales. Thus the fitting was carried
out within two intervals (2 Mpc $\div$ 10 Mpc and 10 Mpc $\div$ 50
Mpc) separately with $\Delta r=1$ Mpc.

For generation of the random catalogue the sky area covered by our
sample was divided into $3^{\circ}\times3^{\circ}$ parts and filled
with the same number (43024 objects) of random points (random
$\alpha$, $\delta$, $z$), preserving the number of objects in each
part and the redshift distribution of initial catalogue. The
absolute magnitudes of the objects ($M_{g}$) in random catalogue
were taken from the initial one and permutated in random way,
preserving redshift dependence of $M_{g}$. One hundred of such
random samples were generated and $\Delta N^{\ast}(r)$ was
calculated for each sample. Then the mean values $\langle\Delta
N^{\ast}(r)\rangle_{100}$ were used instead of $\Delta N^{\ast}(r)$
for fitting with (\ref{fit}).

\begin{table}[!t]\centering
\begin{minipage}[t]{.90\linewidth}
\centering \caption{Parameters of the redshift-space correlation function of quasars with different
luminosities on scales 2 Mpc $\div$ 10 Mpc and 10 Mpc $\div$ 50 Mpc}
\medskip
\begin{tabular}{|c|c|c|c|c|c|}
\hline
 $z$ & $M_{g, eff}$ & $\gamma$  & $r_{c}$, $h^{-1}$Mpc & $\gamma$ & $r_{c}$, $h^{-1}$Mpc \\
  &  & ($r\leq10$ Mpc) & ($r\leq10$ Mpc)  & ($r\geq10$ Mpc)  & ($r\geq10$ Mpc) \\
\hline
               & -24.1 & $2.19\pm0.30$ &  $6.20\pm0.85$ &  $1.40\pm0.32$ &  $7.37\pm1.43$ \\
$0.80\div1.10$ & -25.0 & $2.05\pm0.13$ &  $7.85\pm0.57$ & $1.56\pm0.37$ & $6.12\pm1.73$ \\
               & -25.9 & $1.85\pm0.37$ & $7.97\pm1.18$ & $2.04\pm0.58$ & $6.27\pm1.94$ \\
\hline
               & -24.6 & $1.66\pm0.30$ & $7.61\pm2.21$ & $1.18\pm0.10$ & $11.70\pm1.21$ \\
$1.12\div1.35$ & -25.5 & $1.33\pm0.19$ & $8.31\pm1.73$ & $1.60\pm0.28$ & $7.95\pm1.24$ \\
	       & -26.3 & $2.16\pm0.24$ & $7.24\pm1.27$ & $1.41\pm0.33$ & $6.91\pm1.23$\\
\hline
 	       & -25.1 & $1.46\pm0.17$ & $8.17\pm1.32$ & $1.26\pm0.17$ & $10.21\pm1.15$ \\
$1.35\div1.59$ & -25.9 & $1.83\pm0.35$ & $6.65\pm1.28$ & $1.87\pm0.31$ & $8.27\pm1.17$ \\
 	       & -26.8 & $1.65\pm0.10$  & $12.08\pm1.34$ & $2.30\pm0.36$  & $11.04\pm0.97$\\
\hline
               & -25.2 & $1.51\pm0.42$ & $7.98\pm1.86$ & $1.43\pm0.26$ &  $10.17\pm1.96$ \\
$1.59\div1.83$ & -26.1 & $1.91\pm0.21$ & $7.72\pm0.79$ & $1.17\pm0.11$ & $7.75\pm0.91$ \\
	       & -26.9 & $2.36\pm0.52$ & $6.44\pm0.85$ & $1.94\pm0.52$ &  $4.43\pm1.76$ \\
\hline
	       & -25.6 & $1.30\pm0.19$ & $11.93\pm3.01$ & $1.33\pm0.23$ &  $12.06\pm1.74$ \\
$1.83\div2.20$ & -26.5 & $2.34\pm0.39$ & $4.63\pm0.88$ & $1.40\pm0.30$ & $7.94\pm1.51$ \\
 	       & -27.3 & $1.34\pm0.22$ & $13.56\pm4.51$ & $2.27\pm0.50$ & $8.21\pm2.08$\\
\hline
\end{tabular}
\end{minipage}
\end{table}

Here the errors were calculated in the following way. We obtain the
'initial' values of the parameters ($r_{0}$, $\gamma$), find the deviations of
experimental data from the theoretical one, construct random new
'experimental' data using normal distribution and treating the
theoretical values as mean values and obtained deviations as
dispersions. We generated 100 of such sets of 'experimental' data,
found the parameters for each set, and then find the mean
values of these parameters and their rms, which are presented in
the Table 2.

As one can see, we cannot speak about any significant trend from these results. Some variations of the both parameters are present. E.g. for second and forth z intervals on the scales $r>10$ Mpc the correlation length is smaller for brighter quasars. On all the scales for third and fifth z intervals luminosity dependence of the correlation length seems to follow U-shape, which was found by Porciani \& Norberg \cite{porciani2006}. But all these differences along with variations of $\gamma$ are significant only on 1$\sigma$ level. Thus for varification of these results another another technique, which does not require very big samples was proposed in the next section.

\section{A part of close pairs for
quasars with different luminosities}

\indent \indent The second method lies in direct estimation of a part of quasars with given luminosity which reside in environment with larger quasar density. For the initial sample S1  consider a subsample S2 of quasars having the absolute magnitude $M_{g}$ from a given interval $[M_{g,min},M_{g,max}]$. The number of quasars in subsample S2 is
equal to $N_{tot}(M_{g,min}<M_{g}<M_{g,max})$. For any quasar from S2 we are looking for the nearest neighbours from S1 with any $M_{g}$ at the distance less than r. Let $N(r_{NN}<r, M_{g,min}<M_{g}<M_{g,max})$ be a number of quasars from S2 with $[M_{g,min},M_{g,max}]$ having the distance to the nearest neighbour from S2 less then r. Consider the following function
\begin{equation}\label{f}
    f(r,M_{g})=\frac{N(r_{NN}<r,
    M_{g,min}<M_{g}<M_{g,max})}{N_{tot}(M_{g,min}<M_{g}<M_{g,max})}.
\end{equation}
If we fix the value of r, this function is an estimate of the portion of quasars with given luminosity, having the nearest neighbour distance less r, that could be an estimate of the portion of quasars with given luminosity reside in denser environment. Generally speaking, close pairs of quasars would not necessary reside in clusters. And the same function for the fifth nearest neighbour could be better estimate, but the sample is not large enough for this.

In our case initial samples (S1) are our five samples of quasars from different redshift intervals. In Table 1 $N$ is a number of objects in S1, $N_{tot}$ is a number of objects in subsamples S2.

Note that the function $f(r)$ is a complement to $g(r)$ introduced by White in \cite{white} as
\begin{equation*}\label{g}
    g(r)=\frac{N(r_{NN}>r)}{N_{tot}}.
\end{equation*}
It is readily seen that $f(r)=1-g(r)$. The function $g(r)$ is an estimate of the probability that the volume of radius r is empty. This is a complementary statistic to the correlation functions, because it depends on the correlation functions of all orders in an entirely symmetric way \cite{white}.

If there was no luminosity dependence of the quasar clustering, this function would not depend on absolute magnitude at all. For verification of this statement 100 artificial samples were generated for each from 5 samples in the following way: the absolute magnitudes in the catalogue were permutated in random way, preserving right ascensions, declinations and redshifts on their places. This means that the spatial distribution of quasars (clustering) remains the same, but any luminosity dependence has to disappear.

Results are shown in the Fig.1-5, where only curves for bright and faint quasars are shown for clearness. Open circles denote faint quasars, filled - bright ones. The same curves for permutated samples coincide, thus they are denoted with one curve with open triangles (only in the right parts of the plots). Note, that the errors shown on the plots are statistical ones for initial samples and rms for artificial samples. On the scales less than the clustering scales one can see that the curves for bright quasars lie higher than for faint ones. But the difference between the values $f(r)$ for quasars with different luminosities is statistically insignificant. Only for $1.35<z<1.59$ and $1.59<z<1.83$ this difference is more then 1$\sigma$ on scales of about $r_{c}$. Then the curves intersect on the scales of about half mean nearest neighbour distance. On larger scales the curves for faint quasars lie higher. And on the scale which could be some characteristic scale of the large-scale structure of the Universe these curves coincide. This scale increases with the redshift.

\section{Discussion}

\indent \indent As we can see from the Table 2 to obtain any reliable results with the first method one need much larger samples, which are unavailable for today. Moreover the correlation length is the spatial scale of clustering on the one hand and a measure of the clustering amplitude on the other hand. Thus it is not easy to interpret the first technique results unambiguously. That is why the second method seems to be better for this purpose because it is more
direct and does not require such large samples.

From the results of the second method we can say that brighter quasars reside in closer pairs than faint ones. Note, that such splitting of f(r) curves on larges scales could also be the consequence of the different redshift distributions of the quasars with different luminosities. But we know, that the bright quasars represent higher redshifts and the quasar density decreases with the redshift, thus the inverse effect would be present. Anyway the difference between the mean redshift for subsamples of quasars with different luminosities (see the last column of Table 1) within each redshift interval is negligibly small and cannot affect the results.

Anyway this problem requires further investigations and improvement of the methods. It would be interesting to compare for example, the same $f(r)$ function for the fifth nearest neighbour or cross-correlation with galaxies. The last possibility could increase the sample, but this could be applied only for the lower redshift intervals, than those used in the present work.

Summing up the obtained results one can speal about luminosity dependence of the quasar clustering, but this dependence is not strong, which is in agreement with the results by Porciani \& Norberg \cite{porciani2006} and theoretical predictions by Lidz et al. \cite{lidz}. But even if this dependence is weak, we do not have to
neglect it. Note that when we estimate e.g. the correlation length of the quasars on the low redshift we obtain the mean correlation length averaged over all the quasars with different luminosities. But on the high redshifts the obtained results correspond only to the correlation length for bright quasars and do not reflect the whole picture. That is why possible effects of luminosity dependence of quasar clustering should be taken into account when studying the redshift evolution of it. But even so one should keep in mind that the quasars (as another AGNs) do not reflect the whole distribution of extragalactic objects because quasars are considered to reside in the strongest peaks of the matter density. This fact could explain larger values of the correlation function for quasars than for galaxies (see \cite{blake}, \cite{gonzalez} for comparision).

\section*{Acknowledgments}
\indent \indent The author would like to thank Prof.~V.I.~Zhdanov for his help with the theoretical part of the work and useful discussions and also to both referees for their useful comments and advices. This work has been supported in part by the Cosmomicrophysics program of the  National Academy of Sciences and Space Agency of Ukraine.

\begin{figure}[p]
\begin{minipage}[b]{.95\linewidth}
\centering\epsfig{figure=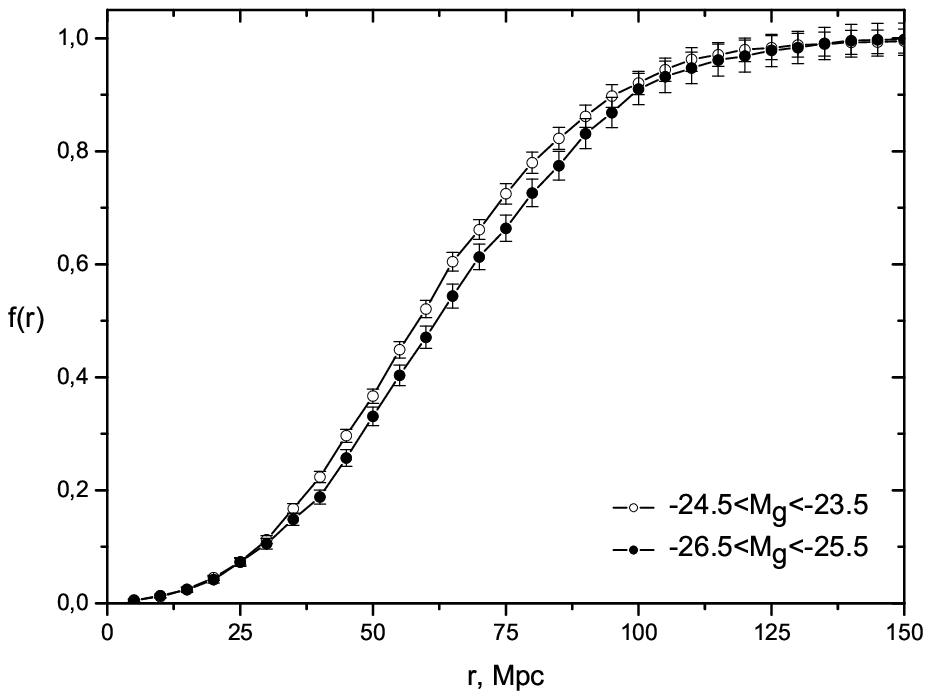,width=.45\linewidth} \hfill
\epsfig{figure=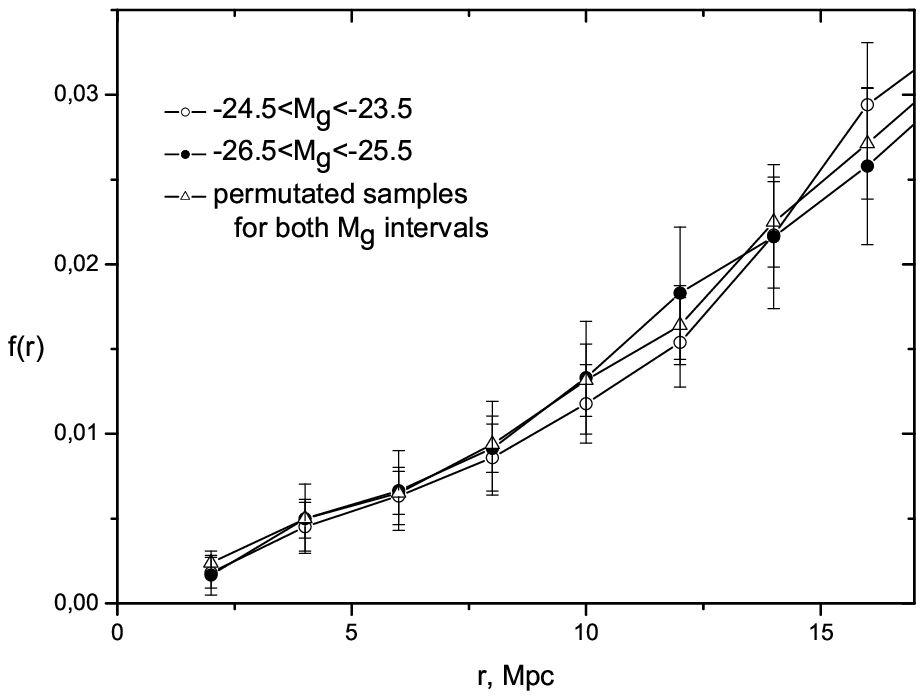,width=.45\linewidth} \caption{$f(r)$ for
$0.80<z<1.10$. Left: for all scales. Right: for small scales.}
\end{minipage}
\begin{minipage}[b]{.95\linewidth}
\centering\epsfig{figure=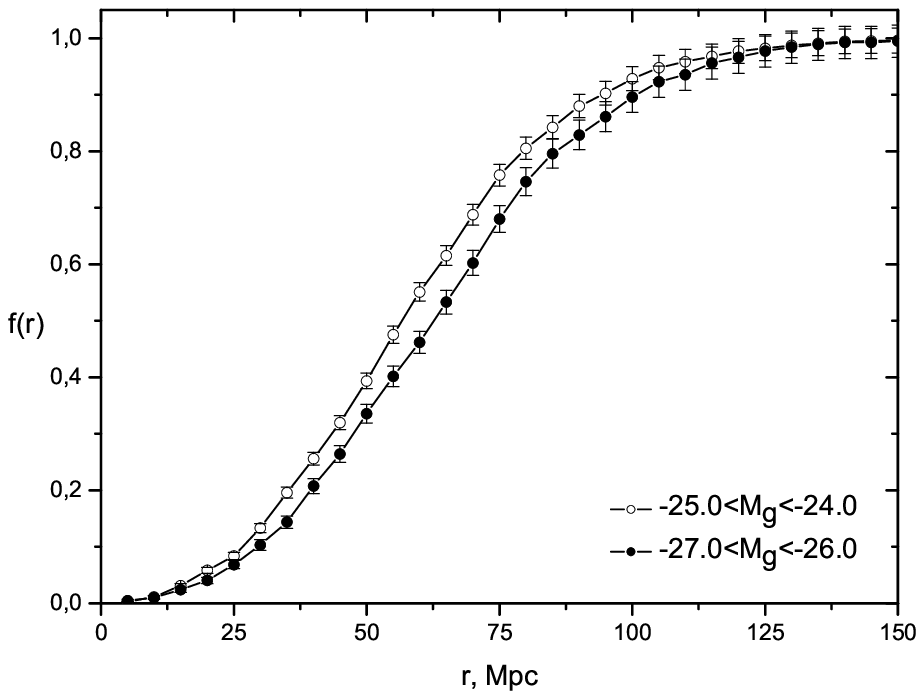,width=.45\linewidth} \hfill
\epsfig{figure=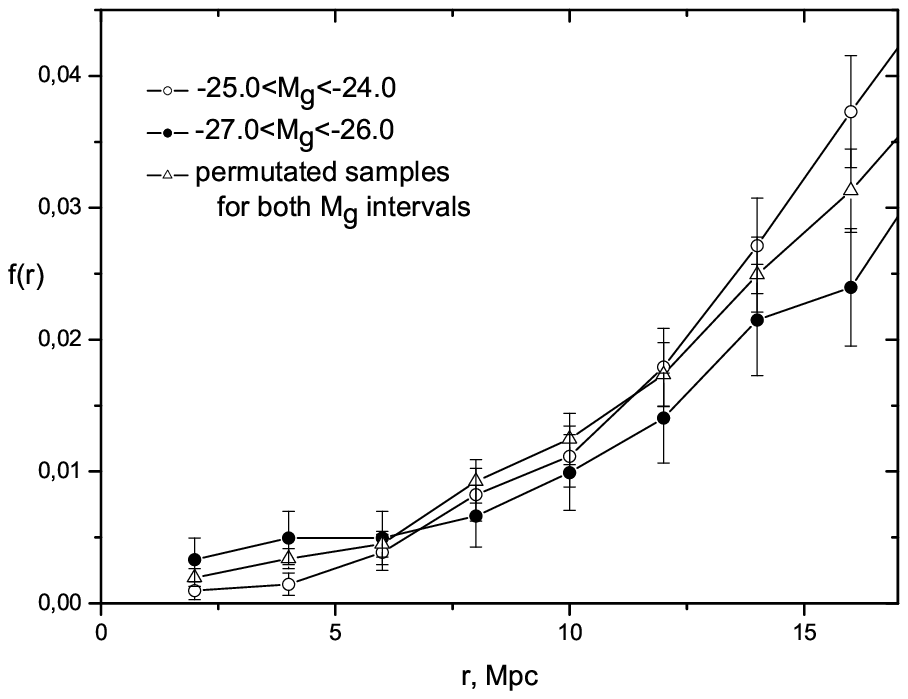,width=.45\linewidth} \caption{$f(r)$ for
$1.10<z<1.35$. Left: for all scales. Right: for small scales.}
\end{minipage}
\begin{minipage}[b]{.95\linewidth}
\centering\epsfig{figure=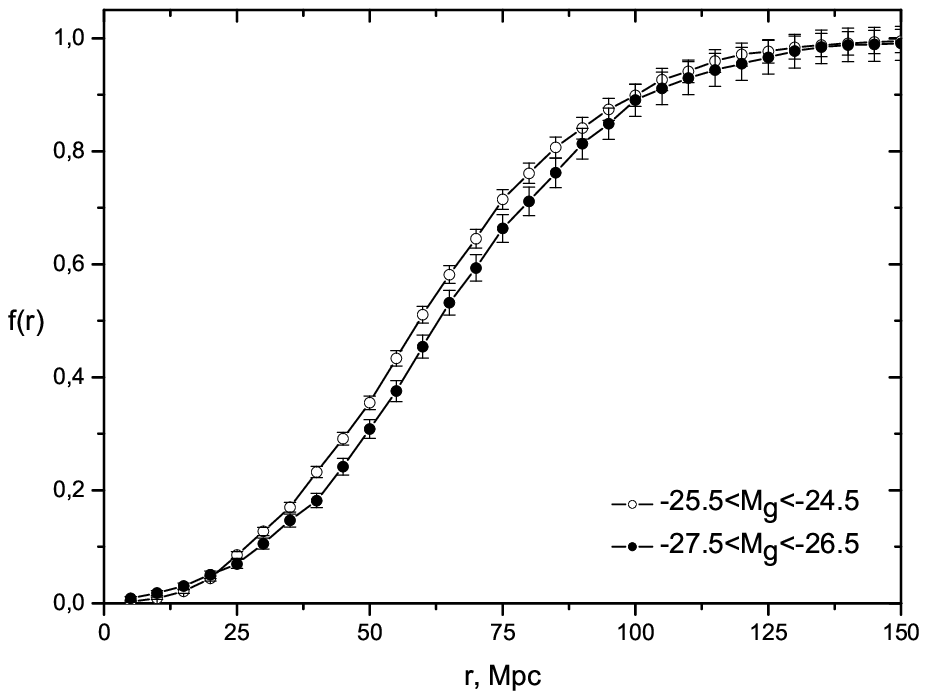,width=.45\linewidth} \hfill
\epsfig{figure=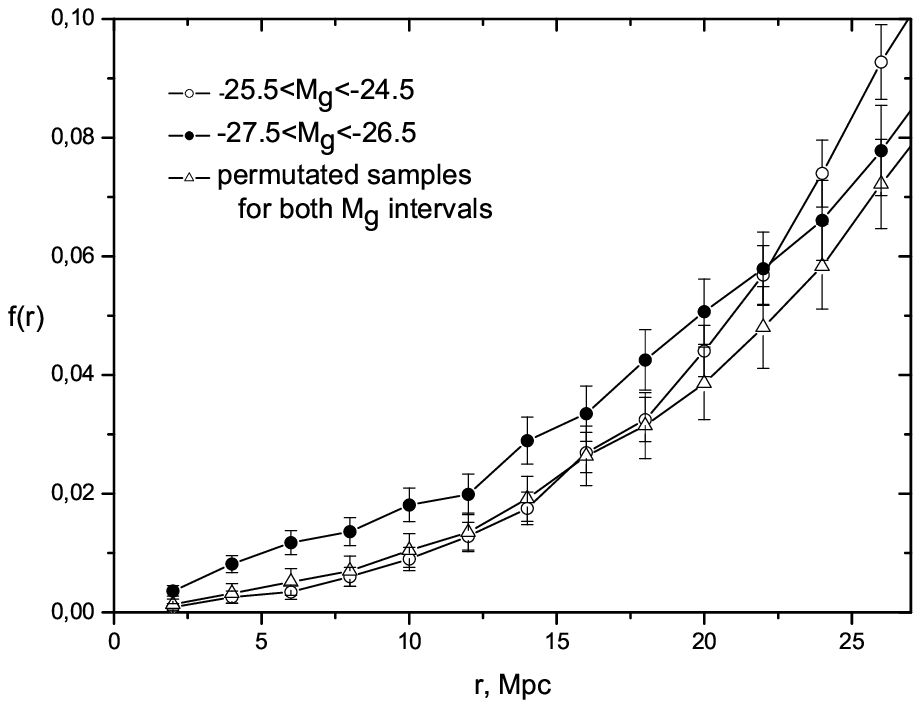,width=.45\linewidth} \caption{$f(r)$ for
$1.35<z<1.59$. Left: for all scales. Right: for small scales.}
\end{minipage}
\end{figure}

\begin{figure}[p]
\begin{minipage}[t]{.95\linewidth}
\centering\epsfig{figure=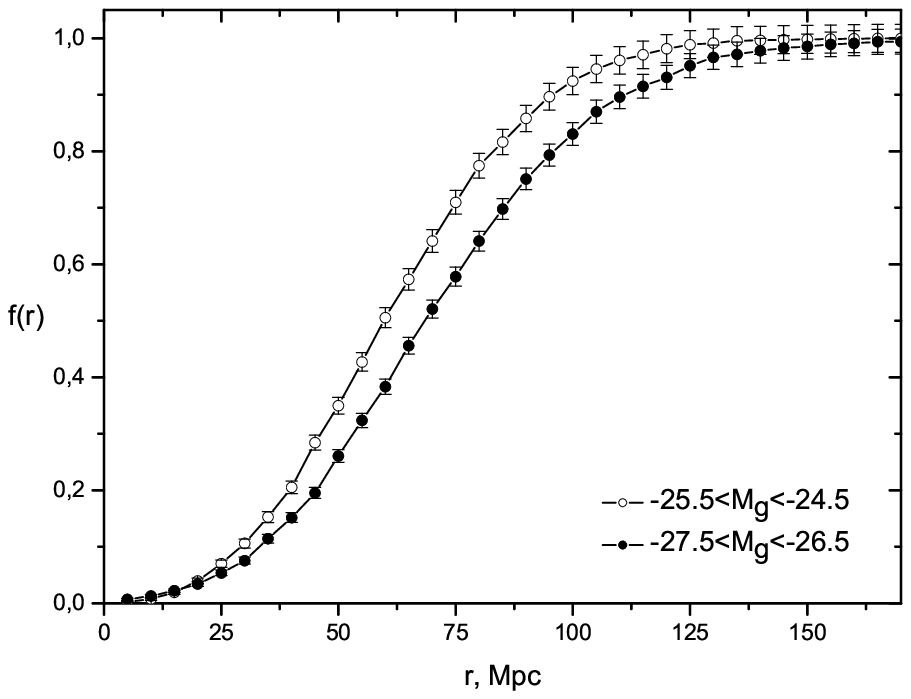,width=.45\linewidth} \hfill
\epsfig{figure=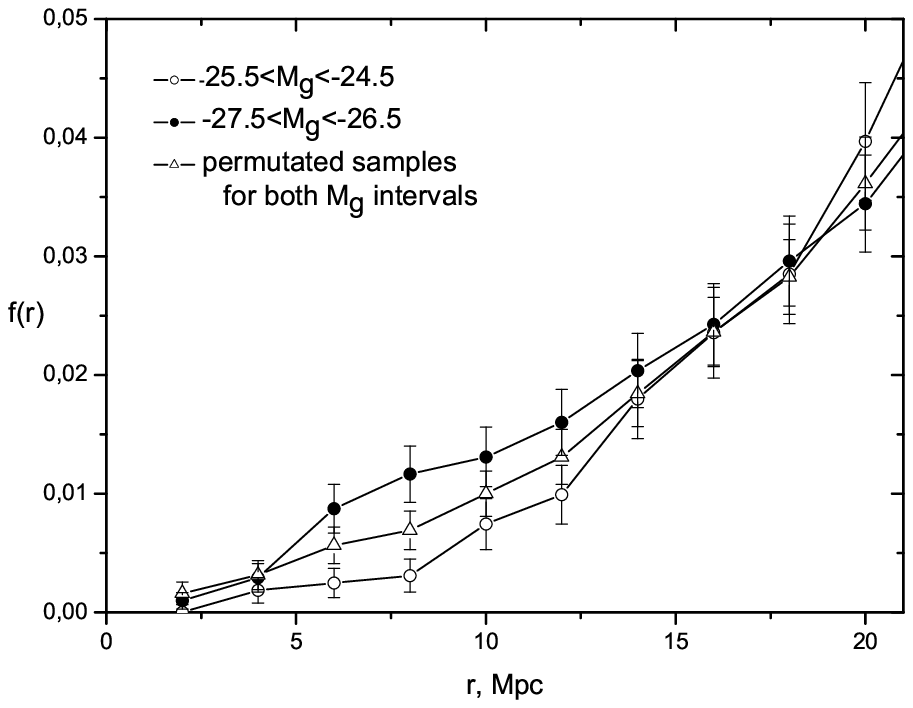,width=.45\linewidth} \caption{$f(r)$ for
$1.59<z<1.83$. Left: for all scales. Right: for small scales.}
\end{minipage}
\begin{minipage}[t]{.95\linewidth}
\centering\epsfig{figure=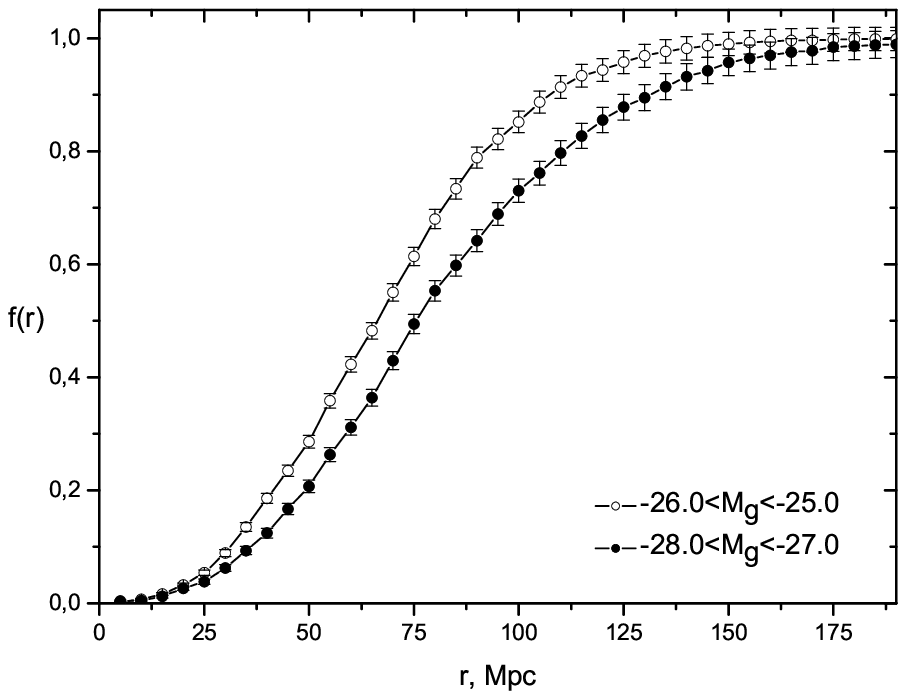,width=.45\linewidth} \hfill
\epsfig{figure=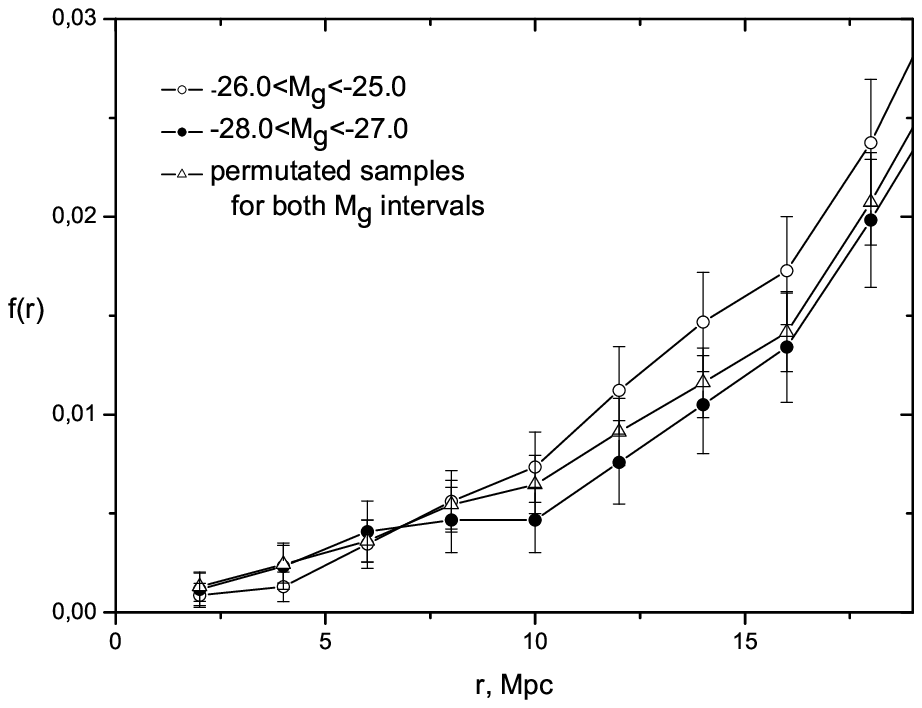,width=.45\linewidth} \caption{$f(r)$ for
$1.83<z<2.20$. Left: for all scales. Right: for small scales.}
\end{minipage}
\end{figure}

%
%
%

\end{document}